\documentclass[aps,prb,twocolumn,floatfix,superscriptaddress,showpacs,10pt,a4paper]{revtex4-1}
\usepackage{graphicx,amsmath,amssymb,amsfonts,sidecap}
\usepackage[usenames,dvipsnames]{color}
\usepackage{bm}
\usepackage{ulem}
\usepackage{placeins}
\usepackage[mathscr]{eucal}

\makeatletter
\newcommand*{\rom}[1]{\expandafter\@slowromancap\romannumeral #1@}
\makeatother

\begin{document}
 
\title{Majorana fermions  in Ge/Si hole nanowires }
\date{\today}
\author{Franziska Maier}
\affiliation{Department of Physics, University of Basel, Klingelbergstrasse 82, CH-4056 Basel, Switzerland}
\author{Jelena Klinovaja}
\affiliation{Department of Physics, Harvard University, Cambridge, Massachusetts 02138, USA}
\author{Daniel Loss}
\affiliation{Department of Physics, University of Basel, Klingelbergstrasse 82, CH-4056 Basel, Switzerland}

\begin{abstract}
We consider Ge/Si core/shell nanowires with hole states coupled to an $s$-wave superconductor in the presence of electric and magnetic fields. We employ a microscopic model that takes into account material-specific details of the band structure such as strong and  electrically tunable Rashba-type spin-orbit interaction and $g$ factor anisotropy for the holes. In addition, the proximity-induced superconductivity Hamiltonian is derived starting from a microscopic model. In the topological phase, the nanowires host Majorana fermions with localization lengths that depend strongly on both the magnetic and electric fields. We identify the optimal regime  in terms of the directions and magnitudes of the fields in which the Majorana fermions are the most localized at the nanowire ends.
In short nanowires, the Majorana fermions hybridize and form a subgap fermion whose energy is split away from zero and oscillates as a function of the applied fields. The period of these oscillations could be used to measure the dependence of the spin-orbit interaction on the applied electric field and the $g$ factor anisotropy.
\end{abstract}

\maketitle

\section{Introduction}
Quasiparticles with non-Abelian statistics are considered as auspicious candidates for topological quantum computing.\cite{Nayak2008} Among these, Majorana fermions (MFs), particles that are their own antiparticles,
received a large amount of attention during the last years. \cite{Alicea2012,Beenakker2013} MFs are predicted to occur in different systems such as fractional quantum Hall systems,\cite{Read2000,Nayak2008} topological insulators,\cite{Fu2008,Sato2009,Tanaka2009,Sasaki2011,Chevallier2012,Sticlet2012,Klinovaja2014} nanowires with strong Rashba\cite{Lutchyn2010, Oreg2010, Alicea2010, Mao2012} or synthetic \cite{Kjaergaard2012,Klinovaja2012a} spin-orbit interaction (SOI), $p$-wave superconductors,\cite{Potter2011} Ruderman-Kittel-Kasuya-Yosida (RKKY) systems,\cite{Klinovaja2013,Braunecker2013,Vazifeh2013}  and graphenelike systems.\cite{Klinovaja2012b,Klinovaja2012c,Klinovaja2013b,Dutreix2013,Klinovaja2013c} 
Recent experiments on MFs \cite{Mourik2012, Das2012,Deng2012, Rokhinson2012,Williams2012,Churchill2013} were performed in Rashba nanowires (NWs) since this type of setup is relatively easy to realize. 
The majority of these experiments use InSb or InAs NWs because they are presumed to have strong SOI and large $g$ factors, which are necessary prerequisites for the emergence of Majorana bound states in such wires.\cite{Lutchyn2010, Oreg2010} However, the direct measurement of the SOI strength in one-dimensional NWs is a challenging task\cite{Rainis2014} and has not yet been performed in the materials mentioned above. In this work we focus on a promising alternative, Ge/Si core/shell NWs carrying holes in the Ge core, in which an exceptionally strong electric-field-highly tunable Rashba SOI is expected \cite{Kloeffel2011} and in which the first signatures of a strong SOI were identified experimentally.\cite{Hao2010} 

Ge/Si core/shell NWs, cylindrical NWs with a Ge core and Si shell, attracted a lot of attention recently. \cite{Lauhon2002,Lu2005,Xiang2006,Xiang2006a,Hu2007,Yan2011,Nah2012,Hu2012}
These NWs can be grown with high precision, i.e., with core diameters between 5 and 100 nm and shell thicknesses between 1 and 10 nm. 
Due to the large valence-band offset between Ge and Si, a one-dimensional (1D) hole gas forms in the core of the NW.\cite{Lu2005,Park2010}
The $p$-type symmetry of the hole Bloch states gives rise to a total angular momentum $J=3/2$, which results in an unusually large, and electrically tunable Rashba-type SOI.\cite{Kloeffel2011}  Furthermore, the holes show high mobilities,\cite{Xiang2006,Nah2012} long mean free paths,\cite{Lu2005} and Coulomb interaction strongly influences their properties.\cite{Maier2014}
Longitudinal confinement in these NWs results in tunable single and double QDs \cite{Hu2007} with anisotropic and confinement-dependent $g$ factors,\cite{Roddaro2007,Roddaro2008} in long relaxation \cite{Hu2012} and coherence times \cite{Higginbotham2014b} as well as in short SOI lengths. \cite{Higginbotham2014a}
Moreover, strongly anisotropic tunable $g$ factors and long spin phonon relaxation times \cite{Maier2013} were predicted as well as the usability for quantum information processing based on hole spin qubits. \cite{Kloeffel2013} 
 Note that strongly anisotropic and electrically tunable $g$ factors were also observed in SiGe nanocrystals. \cite{Ares2013,Ares2013a} 
Most importantly for our work, externally applied strong magnetic fields allow one to access a helical regime,\cite{Kloeffel2011} which in combination with the experimentally demonstrated proximity-induced $s$-wave superconductivity\cite{Xiang2006a} makes Ge/Si core/shell NWs promising candidates to generate MFs.

In the present work, we explore the properties of MFs in Ge/Si core/shell NWs starting from a microscopic model \cite{Kloeffel2011} that captures the specific NW characteristics such as $g$ factor anisotropy and the dependence of the induced Rashba SOI on the direction and magnitude of the electric field. We extend this microscopic model to account on the same level for a proximity-induced superconductivity, which couples hole states with opposite orbital and angular momentum. 
We especially focus on the tunability of the SOI that allows us to access the regimes of strong and weak SOI independently of the applied magnetic field
and analyze the localization lengths of the MF wave functions in the NW regarding their dependence on magnitude and direction of the applied electric and magnetic fields.
The shortest localization lengths can be expected when the fields are tuned to intermediate magnitudes. 
Due to the $g$ factor anisotropy we predict the shortest localization lengths for the magnetic field pointing perpendicular to the NW axis.
In a NW of finite length, the MFs localized at two NW ends overlap\cite{Prada2012} and hybridize into an ordinary fermion which is, generally, at nonzero energy.
This energy demonstrates an oscillatory behavior\cite{DasSarma2012, Rainis2013} as a function of the applied electric and magnetic fields that might be used to determine the coupling constants to the electromagnetic field.
Finally, we mention in passing that while we focus here on  Ge/Si core-shell nanowires, we expect our analysis also to apply (at least qualitatively) to similar structures such as Ge hat-shape nanowires grown on Si.~\cite{Zhang2012}

The outline of the paper is as follows. 
In Sec.~\ref{sec:Hamiltonian}, we introduce the effective microscopic 1D model and derive the proximity-induced superconducting coupling of NW hole states. In Sec.~\ref{sec:MajWavefuncts}, we determine the localization lengths of MFs for a semi-infinite NW in both the strong and the weak SOI regime and identify field configurations for which the shortest localization lengths can be expected. Last,  we focus on the energies of hybridized MFs in finite NWs in Sec.~\ref{sec:Energyoscillations}. 
We present our conclusions in Sec.~\ref{sec:conclusion}. 
Technical details are deferred to the Appendixes. 
%
%
%
%
%

\section{ Nanowire Hamiltonian for holes \label{sec:Hamiltonian}}
%
%
%
%
%
In this section, we describe the geometry of the system and describe the directions of the applied fields for which MFs can be expected. 
We introduce the microscopic model\cite{Kloeffel2011} describing holes confined to the core of Ge/Si core/shell NWs and derive an effective lowest-energy subband Hamiltonian and the associated spectrum. 
Next, we employ a superconductivity pairing Hamiltonian introduced for holes close to the valence-band edge in bulk material\cite{Mao2012} and derive the corresponding 1D Hamiltonian within the framework of the microscopic model and project this on the subspace of the lowest-energy subband Hamiltonian.
%
%
%
%
%
\subsection{Setup \label{sec:SetupSketch}}
\begin{figure}[tbp]
\includegraphics[width = \columnwidth]{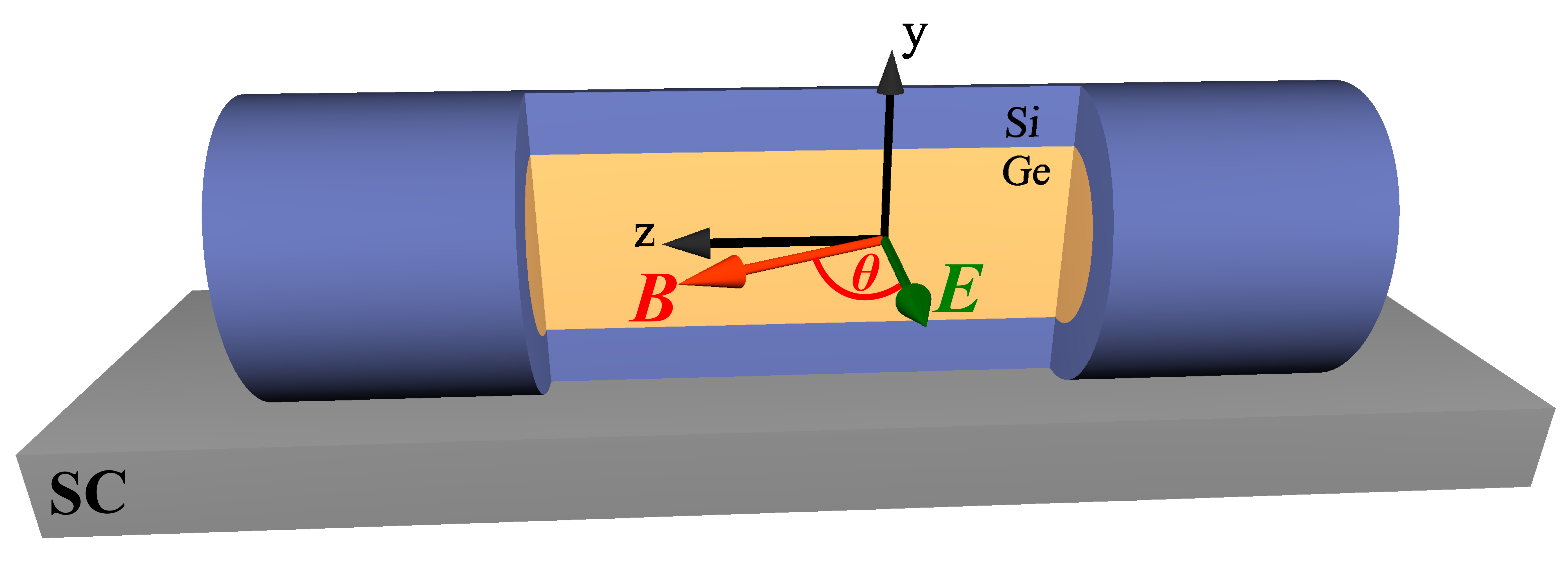}
\caption{Sketch of a  Ge/Si core/shell NW (cylinder) placed on top of an $s$-wave superconductor (SC) to induce proximity pairing of the holes in the NW core. The NW axis is chosen to point along the $z$ axis.  The applied  electric field $\bm{E} = (E_x,0,0)$ is parallel to the $x$ axis, while the applied magnetic field $\bm{B} = (B_x,0,B_z) \equiv B(\cos\theta,0,\sin\theta)$ is in the $xz$ plane. 
\label{fig:wiresetup}}
\end{figure}

Throughout this work, we consider holes confined to the core of a Ge/Si core/shell NW with core (shell) radius $R$ $(R_s)$ that is positioned on top of an $s$-wave superconductor as sketched in Fig.~\ref{fig:wiresetup}. The NW axis is assumed to point along the $z$ axis.
We restrict ourselves to field configurations in which the electric field $\bm{E}= (E_x,0, 0)$  points perpendicular to the NW axis and is parallel to the surface of the superconductor  and in which the magnetic field $\bm{B} = (B_x,0,B_z)\equiv B(\cos\theta,0,\sin\theta)$ is confined to the plane spanned by $\bm{E}$ and the NW axis. In this case, the SOI vector, generated by the applied electric field $\bm{E}$, points along the $y$ axis and is always perpendicular to $\bm{B}$.  As shown before,\cite{Lutchyn2010,Oreg2010} such a configuration is optimal for generating MFs in NWs.
We note that we focus here on the case where the SOI (and the electric field $\bm{E}$) is uniform along the NW.  For the effects of a nonuniform SOI we refer to Ref.~[\onlinecite{Klinovaja2014a}]. 

%
%
%
%
%
\subsection{Microscopic Hamiltonian \label{sec:ModelKloeffel}}
The two hole bands closest to the valence-band edge in a Ge/Si core/shell NW are  described by an effective $4\times4$ Hamiltonian \cite{Kloeffel2011}
\begin{equation}
H = \sum_{ij} \int dz\, \Psi_i^{\dagger}(z) (\mathscr{H}_{ij} + \mu \delta_{ij}) \Psi_j(z),\label{eq:Ham4x4}
\end{equation}
with Hamiltonian density 
\begin{equation}
\mathscr{H}=\mathscr{H}_{\text{LK}} + \mathscr{H}_\text{strain} +\mathscr{H}_{\text{DR}}+\mathscr{H}_{\text{R}}+ \mathscr{H}_{B,Z}+\mathscr{H}_{B,\text{orb}}, \label{eq:Ham4x4_split}
\end{equation}
given in the basis $\{\Psi_{g_+}(z), \Psi_{g_-}(z),\Psi_{e_+}(z), \Psi_{e_-}(z)\}$.
Here, $\mu$ is a tunable chemical potential, and $\delta_{ij}$ is the Kronecker $\delta$.
The fermionic annihilation operators $\Psi_{i}(z) = \sum_{k_z} e^{i k_z z} c_{i,k_z}$ can be rewritten in momentum space in terms of $c_{i,k_z}$, which are the fermionic annihilation operators of a hole state $i\in \{g_{\pm},e_{\pm}\}$ with momentum $k_z$ along the NW. The index $g,e$ refers to the ground and excited bands, and the index $\pm$ refers to the pseudospin.

The Luttinger-Kohn and strain Hamiltonian densities are given by 
\begin{equation}
\mathscr{H}_{\text{LK}} + \mathscr{H}_\text{strain} = A_{+}(k_z) + A_{-}(k_z) \tau_z +  C k_z \tau_y \sigma_x,
\end{equation}
where $\tau_i$ and $\sigma_i$ denote the Pauli matrices for band $(g,e)$ and pseudospin index $(+,-)$, respectively. 
Here, $A_{\pm}(k_z, \eta) \equiv \hbar^2 k_z^2(m_g^{-1} \pm m_e^{-1})/4\mbox{ $\pm$ } \Delta/2$, with Planck's constant $\hbar$ and effective masses $m_g\simeq m_0/(\gamma_1+2\gamma_s)$ and $m_e = m_0/(\gamma_1+\gamma_s)$ with $m_0$ denoting the bare electron mass and $\gamma_1$ and $\gamma_s$ representing the Luttinger parameters in spherical approximation. 
For Ge, $\gamma_1=13.35$ and $\gamma_s=5.11$.\cite{Lawaetz1971}
The level splitting is $\Delta \equiv \Delta_{\rm LK} + \Delta_{\rm strain}(\eta)$ with confinement induced $\Delta_{\text{LK}}=0.73 \hbar^2/(m_0 R^2)$ and the strain dependent splitting $\Delta_{\text{strain}}(\eta)\simeq0-30\mbox{ meV}$, where the latter depends on the relative shell thickness $\eta \equiv (R_s - R)/R$. 
The off-diagonal terms, being proportional to the coupling constant $C=7.26 \hbar^2/(m_0 R)$, result directly from the strong SOI at the atomic level.
The direct Rashba SOI, 
\begin{equation}
\mathscr{H}_{\text{DR}} = e U E_x \tau_x \sigma_z, \ \ \ \ U=0.15 R,
\end{equation}
originates from the direct dipolar coupling of the hole charge to the applied electric field $E_{x}$. The conventional Rashba SOI reads
\begin{equation}
\mathscr{H}_{\text{R}}= \alpha_R E_{x} [S \tau_x \sigma_z + B_{+}(k_z) + B_{-}(k_z) \tau_z ],
\end{equation}
with $B_{\pm}(k_z) \equiv k_z T \sigma_y/2\pm3 k_z   \sigma_y/8$, where $T=0.98$, $S=0.36/R$, and $\alpha_R=-0.4 \mbox{ nm}^2 e$ with elementary charge $e$.
The parameters $S$ and $U$ of the direct and conventional SOI, respectively, are related by $e U/(\alpha_R S) \simeq -1.1 R^2\mathrm{nm}^{-2}$, hence $\mathscr{H}_{\text{DR}}$ dominates $\mathscr{H}_{\text{R}}$ by one to two orders of magnitude for  $R=5-10\mbox{ nm}$.~\cite{Kloeffel2011}

Finally, we include the effect of an applied magnetic field $\bm{B} = (B_x,0,B_z)=B(\cos\theta,0,\sin\theta)$ by introducing the Hamiltonian densities 
\begin{align}
\mathscr{H}_{B,Z} &= [C_{+}+C_{-}\tau_z]\sigma_z +[D_{+}+D_{-}\tau_z]\ \sigma_x,\\
\mathscr{H}_{B,\text{orb}} &= F_z \tau_x \sigma_y+ F_{x}\tau_y .
\end{align}
Here, $C_{\pm} = \mu_B B_z (F\pm G)/2$, $D_{\pm} = \mu_B B_x (K\pm M)/2$, $F_z = \mu_B B_z D k_z$ and $F_x = \mu_B B_x L_B k_z$ with $F = 1.56$, $G = -0.06$, $K = 2.89$, $M = 2.56$, $D = 2.38 R$, and $L_B = 8.04 R$. \cite{Kloeffel2011}
%
%
%
%
%
%
\subsection{Low-energy $2\times2$ Hamiltonian \label{sec:derivationLEHamiltonian}}
In this subsection, we derive an effective lowest-energy subband Hamiltonian for the holes by effectively decoupling the $g_{\pm}$ and $e_{\pm}$ bands introduced above in Sec.~\ref{sec:ModelKloeffel}. 
To achieve this, we perform a Schrieffer-Wolff transformation\cite{Winkler2003,Bravyi2011} (SWT) which block diagonalizes the Hamiltonian and subsequently allows one to truncate the lowest-energy subspace.
In general, a SWT is given by a transformation of the form $H \rightarrow \tilde{H} = e^{-S}He^{S}$, where $S$ is an anti-Hermitian operator ($S^{\dagger} = -S$). 
However, we utilize the SWT in a perturbative manner and begin by subdividing the Hamiltonian density $\mathscr{H}$ into a leading order term $\mathscr{H}_0 = A_{+}(0) + A_{-}(0) \tau_z$ and a perturbation $\mathscr{H}' = \mathscr{H} -\mathscr{H}_0$.
This choice is justified since the strain induced splitting of the $g_{\pm}$ and $e_{\pm}$ subspaces is by far the largest energy scale present in the system. 
The perturbing term is further divided into a diagonal ($\mathscr{H}_d$) and off-diagonal part ($\mathscr{H}_{od}$) with respect to the two considered subspaces $g_{\pm}$ and $e_{\pm}$, $\mathscr{H}' = \mathscr{H}_d+  \mathscr{H}_{od}$. 
Next, we construct the operator $S$ such that the SWT rotates $\mathscr{H}_{od}$ into an approximately block-diagonal form. 
We also expand $e^{S} \approx 1+ S+ S^2/2$ and then approximate $S$ to lowest order by $S \approx S_1$, where $S_1$ is determined by $[S_1,  \mathscr{H}_0] =  \mathscr{H}_{od}$. As a result, the approximate block-diagonal Hamiltonian density $\tilde{\mathscr{H}}\approx \mathscr{H}_0+ \mathscr{H}_d- [S_1,  \mathscr{H}_{od}]+[S_1,[S_1, \mathscr{H}_0]]/2$ is exact to second order in $(\mathscr{H}_{od})_{ij}/\Delta\ll1$, where $(\mathscr{H}_{od})_{ij}$ denotes the matrix elements coupling the $g_{\pm}$ and $e_{\pm}$ subspaces.
This corresponds to conditions restricting the magnitudes of the applied fields: $C k_z/\Delta\ll1$, $e U E_x/\Delta\ll1$, $\mu_B B_x L k_z/\Delta\ll1$, and $\mu_B B_z D k_z/\Delta\ll1$. 
After truncating, the effective lowest-energy Hamiltonian is given by $\tilde{H}_{g'}=\sum_{i,j=g_{\pm}'}\int\mathrm{d}z \Psi_i (\tilde{\mathscr{H}}_{g'})_{ij}\Psi_j$, with density
\begin{align}
&\tilde{\mathscr{H}}_{g'} = \nonumber\\
&\left(\begin{array}{cc}
\frac{\hbar^2}{2 m_{\text{eff}}}k_z^2 - \mu + \mu_B B_z g_{z}  & \mu_B B_x g_{x}- i E_x k_z  \alpha_{\text{eff}}\\
\mu_B B_x g_{x} + i E_x k_z  \alpha_{\text{eff}}& \frac{\hbar^2}{2 m_{\text{eff}}}k_z^2 - \mu -  \mu_B B_z g_{z}
\end{array}\right)  \label{eq:HeffKloeffel}
\end{align}
in the associated low-energy basis $\{\Psi_{g'_+},\Psi_{g'_-}\}$. 
The new annihilation operators are linear combinations of the original operators introduced below 
Eq.~(\ref{eq:Ham4x4_split}), where the associated admixing coefficients depend strongly on the NW parameters $R$ and $\Delta$ and on the magnitude and direction of $\bm{E}$ and $\bm{B}$.

In Eq.~(\ref{eq:HeffKloeffel}), we identify an effective kinetic term $\propto m_{\text{eff}}^{-1}$, an effective SOI term $\propto \alpha_{\text{eff}}$, and two terms $\propto g_x, g_z$ describing the effective coupling to the magnetic field, 
\begin{align} 
\alpha_{\text{eff}} =&\, T \alpha + \frac{2}{\Delta}C (e U + S \alpha), \ \frac{\hbar^2}{2 m_{\text{eff}}}  \approx \frac{\hbar^2}{2 m_g}-\frac{C^2}{\Delta},\label{eq:meff}\\
g_{x} =& K-\frac{2}{\Delta}L_B C k_z^2,  \ \ \ \ \ \ \ \  g_{z} = G-\frac{2}{\Delta}D C k_z^2. \label{eq:gzeff}
\end{align}
We note that $m_{\text{eff}}$ has an additional weak dependence on $\bm{B}$ which is neglected here. Furthermore, we see that the effective $g$ factors $g_i = g_{i0}+ g_{i2}k_z^2$ ($i=x,z$) differ strongly in magnitude, which leads to anisotropy, and, in addition, they depend on the momentum $k_z$.

The Hamiltonian $\tilde{H}_{g'}$ describes the lowest-energy subbands $g_{\pm}'$, where all coupling terms to the higher bands are taken into account by introducing effective $g$ factors and SOI coupling.
We note that the Hamiltonian density $\tilde{\mathscr{H}}_{g'}$ resembles closely the Hamiltonian densities introduced  in other works\cite{Oreg2010, Lutchyn2010, Klinovaja2012} to describe electrons in  Rashba SOI NWs in the presence of a magnetic field. However, the dependence on the direction and strength of $\bf E$ and $\bf B$  is much more involved in the case of Ge/Si core/shell NWs.

The spectrum of $\tilde{H}_{g'}$ is given by
\begin{equation}
E_{u,d}(k_z) = \frac{\hbar^2 k_z^2}{2 m_{\text{eff}}} \pm \sqrt{E_x^2 \alpha_{\text{eff}}^2 k_z^2 + \Delta_Z^2},
\end{equation}
where $\Delta_Z^2 = \mu_B^2(B_x^2 g_x^2+ B_z^2 g_z^2)$ and the index $u$ ($d$) marks the upper (lower) energy band. 
\begin{figure}[tbp]
\includegraphics[width = \columnwidth]{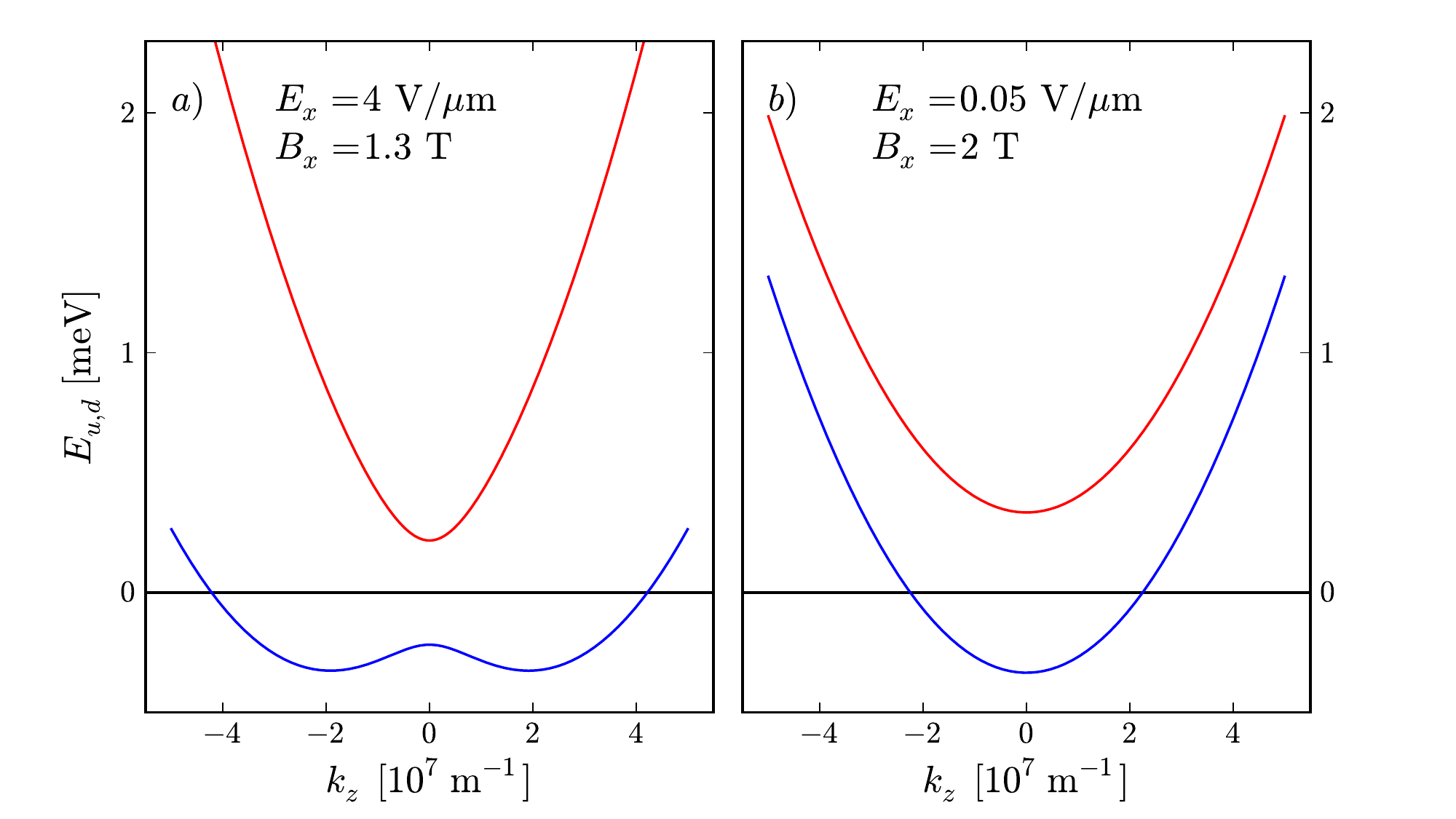}
\caption{The lowest-energy spectra $E_{u}(k_z)$ (red) and $E_{d}(k_z)$ (blue) as functions of the momentum $k_z$ (a) in the strong SOI regime and (b) in the weak SOI regime. The respective magnitudes of the applied fields are given as insets. 
The magnetic field $\bm{B}$ is chosen along the $x$ axis (like the $\bm{E}$ field). The used NW parameters are $R=7.5\mbox{ nm}$ and $\Delta = 23\mbox{ meV}$. 
\label{fig:LE-spectrum}}
\end{figure}
The Fermi wave vector $k_F$ is found from the condition $E_{d}(k_z)=0$ and is given by
\begin{equation}
k_F = \pm\sqrt{2 \tilde{k}_{so}^2+ \sqrt{4 \tilde{k}_{so}^4 + \tilde{k}_{Z}^4}} ,\label{eq:kFermi_general}
\end{equation}
with components
\begin{align}
\tilde{k}_{so}^2 =& \frac{    E_x^2 \alpha_{\text{eff}}^2 -4 (C/\Delta) \left[B_z^2 D G+B_x^2 K L_B\right] \mu_B^2 }{ \hbar ^4/m_{\text{eff}}^2-16 (C/\Delta)^2 \left[B_z^2 D^2+B_x^2 L_B^2\right]  \mu_B^2}, \\
\tilde{k}_{Z}^4 =&  \frac{4\left[B_z^2 G^2+B_x^2 K^2\right]  \mu_B^2}{ \hbar ^4/m_{\text{eff}}^2-16 (C/\Delta)^2 \left[B_z^2 D^2+B_x^2 L_B^2\right]  \mu_B^2}.
\end{align}
To obtain this result, we have taken into account the full $k_z$ dependence of the $g$ factors, hence we cannot use the effective parameters $g_x$ and $g_z$ as they themselves depend on $k_z$ [see Eq.~(\ref{eq:gzeff})].
Using the definitions above, we introduce the SOI energy $\tilde{\Delta}_{so} = E_x \alpha_{\text{eff}} \tilde{k}_{so}/2$. This allows us to  distinguish between the strong SOI regime ($k_F/\tilde{k}_{so}\approx2$), where the SOI energy dominates over the Zeeman energy,  and the weak SOI regime ($k_F/\tilde{k}_{Z}\approx1$), where the Zeeman energy dominates over the SOI energy. 
In Fig.~\ref{fig:LE-spectrum}, we plot the bands $E_{u,d}(k_z)$ for two sets of finite electric and magnetic fields: one in the strong  [Fig.~\ref{fig:LE-spectrum}(a)] and one in the weak  [Fig.~\ref{fig:LE-spectrum}(b)] SOI regime. 
We compared numerically the exact spectrum with the approximate one for
the configurations considered in Fig.~\ref{fig:LE-spectrum} and found good agreement.
%
%
%
%
%
\subsection{Superconductivity: Pairing Hamiltonian \label{sec:superconductingHam}}
In this section, we derive an effective Hamiltonian describing the proximity induced superconductivity in the lowest subband $g'$. The proximity-induced superconducting pairing gap $\Delta_{sc}^{\text{exp}}\approx235 \mbox{ $\mu$eV}$ was observed experimentally in Ge/Si core/shell NWs.\cite{Xiang2006} Pairing Hamiltonians describing superconductivity for hole states in semiconductors were also discussed before in several theoretical works. \cite{Futterer2011,Mao2012, Moghaddam2014} Here, we start from a general pairing Hamiltonian allowing for coupling between bulk hole states with opposite orbital and angular momentum,\cite{Mao2012}
\begin{align}
H_{SC} =\int d^3r&\left[\Delta_{3/2}\Psi_{3/2}^{\dagger}\Psi_{-3/2}^{\dagger} \right.\nonumber\\
& \left. +\Delta_{1/2} \Psi_{1/2}^{\dagger}\Psi_{-1/2}^{\dagger}+ \text{H.c.}\right],\label{eq:Hsupercond_bulk}
\end{align}
where the fermionic operators $\Psi_{m_j}$ annihilate bulk hole states with angular momentum $j=3/2$ and $m_j = \pm3/2,\pm1/2$ and which are coupled by the respective superconducting pairing potentials $\Delta_{3/2}$ and $\Delta_{1/2}$. 
We assume that $\Delta_{3/2}$ is real but employ $\Delta_{1/2} = |\Delta_{1/2}| e^{i\varphi_{sc}}$ to account for a possible complex superconducting phase. 
We use $H_{SC}$ to derive an effective paring Hamiltonian within the framework of the microscopic model (see Sec.~\ref{sec:ModelKloeffel}) by modifying the procedure outlined in Ref.~[\onlinecite{Kloeffel2011}].
By extending the basis of the microscopic model given below Eq.~(\ref{eq:Ham4x4_split}) accordingly, we derive an effective 1D particle-hole basis. 
\footnote{To avoid confusion between various holes, we note that the hole mentioned here actually denotes the conjugate hole.}
Furthermore, we use the explicit three-dimensional wave functions of the hole states in the NW\cite{Kloeffel2011} and calculate the matrix elements of the effective 1D superconducting Hamiltonian by integrating out the transverse part. 
Finally, we transform the resulting Hamiltonian by the SWT introduced in Sec.~\ref{sec:derivationLEHamiltonian} and truncate the lowest-energy particle-hole subspace with Nambu space representation $\Psi_{ph} = (\Psi_{g_+'},\Psi_{g_-'},\Psi_{g_+'}^{\dagger},\Psi_{g_-'}^{\dagger})$. 
In this representation, the effective lowest-energy subband superconducting pairing Hamiltonian is given by $\tilde{H}_{SC} = \frac{1}{2}\int dz \Psi_{ph}^{\dagger}\tilde{\mathscr{H}}_{SC} \Psi_{ph}$ with
\begin{equation}
\tilde{\mathscr{H}}_{SC} =\left(\begin{array}{cccc}
    0	& 0	& 0	&i \Delta_{sc}\\
    0	& 0	& -i \Delta_{sc}	& 0\\
    0	& i \Delta_{sc}^*	& 0	& 0\\
   -i \Delta_{sc}^*	& 0	& 0	& 0
\end{array}\right), \label{eq:Hamsupercondeff1D}
\end{equation}
where $i \Delta_{sc} = 0.01 \Delta_{3/2}- 0.5 |\Delta_{1/2}| e^{i \varphi_{sc}}$.
We combine $\tilde{\mathscr{H}}_{SC}$ with $\tilde{\mathscr{H}}_{g'}$, where the latter is extended to the particle-hole subspace $\Psi_{ph}$, and obtain an effective Bogoliubov-de Gennes Hamiltonian (explicitly given in Appendix \ref{app:Hamiltonian}).  
The spectrum of this Hamiltonian is given by 
\begin{align}
E^2 &= \left(\frac{\hbar^2 k_z^2}{2 m_{\text{eff}}}\right)^2+E_x^2 k_z^2 \alpha_{\text{eff}}^2+\Delta_Z^2+|\Delta_{sc}|^2\nonumber\\
&\pm2\sqrt{\left(\frac{\hbar^2 k_z^2}{2 m_{\text{eff}}}\right)^2 \left(E_x^2 k_z^2 \alpha_{\text{eff}}^2+\Delta_Z^2\right) + |\Delta_{sc}|^2 \Delta_Z^2 }. \label{eq:spectrumHph}
\end{align}
At $k_z=0$, we find that Eq.~(\ref{eq:spectrumHph}) reduces to $E^2 = (|\Delta_{sc}|\pm \Delta_Z)^2$, and the topological gap is given by 
\begin{equation}
\Delta_{-} = |\Delta_{sc}|-\Delta_Z. 
\end{equation}
The system is in the nontopological phase for $\Delta_{-} >0$ and in the topological phase for $\Delta_{-} <0$.\cite{Oreg2010, Lutchyn2010, Klinovaja2012}

%
%
%
%
%
\section{Tunability of the MF localization length \label{sec:MajWavefuncts}}
Next, we focus on the MF wave functions and associated localization lengths assuming that the two MFs are well localized at the ends of a Ge/Si core/shell NW and do not overlap with each other.
To obtain independent solutions for the MF wave functions at both ends, we simplify the calculations by assuming a semi-infinite NW originating, let's say, at $z=0$.
For topological computational schemes, one generally strives for small localization lengths and, thus, well localized MFs. 
We analyze the tunability of the localization lengths as functions of magnitude and direction of the applied fields $\bm{E}$ and $\bm{B}$ and determine the regime in which the localization lengths are the shortest.
%
%
%
%
%
\subsection{Strong SOI \label{sec:derivationwavefunct_strongSOI}}
\begin{figure}[tbp]
\includegraphics[width=\columnwidth]{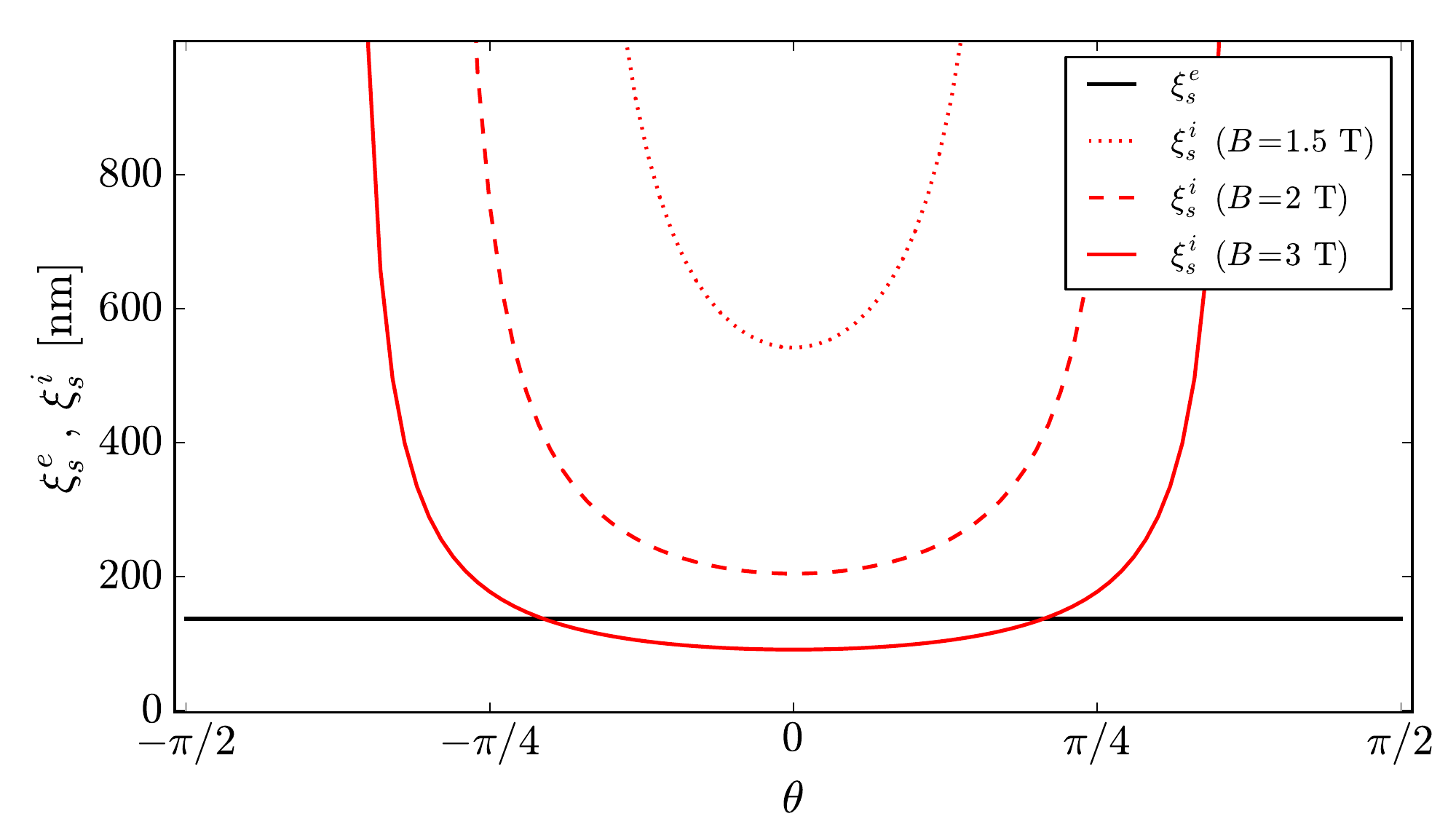}
\caption{Localization lengths $\xi_s^e$ and $\xi_s^i$ as functions of the angle $\theta$ defined by $\bm{B} = B(\cos\theta,0,\sin\theta)$, for  $E_x =4 \mbox{ V/$\mu$m}$ and for different magnitudes of $B$. 
As soon as $\theta$ deviates from $0$, $\xi_s^i$ increases until its value diverges when the topological gap closes. 
Furthermore, increasing $B$ above a threshold value given by $|\Delta_{-}|=|\Delta_{sc}|$, i.e.\ $\Delta_Z^0\geq 2|\Delta_{sc}|$, results in $\xi_s^i<\xi_s^e$ for a certain range of $\theta$. 
Since $\xi_s^e$ is independent of $\bm{B}$, its value appears as a constant in the plot. 
We use the NW parameters $R=7.5\mbox{ nm}$ and $\Delta = 23\mbox{ meV}$ and assume a superconductivity pairing potential of $\Delta_{sc} = 200 \mbox{ $\mu$eV}$. 
\label{fig:localizationlengths_strongSOI_Bangledep}}
\end{figure}

First, we focus on the strong SOI regime, where $\Delta_{so}\gg\Delta_Z$ with $\Delta_{so} =  \alpha_{\text{eff}} E_x k_{so}/2$ and $k_{so}=m_{\text{eff}} \alpha_{\text{eff}}E_x/\hbar^2$. 
Here, the Fermi wave number is given by $k_F^s = 2 k_{so}$, and the associated Fermi velocity is $v_F^s = \alpha_{\text{eff}} E_x/\hbar$. 
In this regime, the main effect of the applied magnetic field is the opening of a gap at $k_z=0$, thus we are allowed to drop the $k_z$ dependence of the $g$ factors [see Eqs.~(\ref{eq:gzeff})]
 and to use the following approximation for the Zeeman splitting $\Delta_Z \approx \Delta_Z^0 e^{i \vartheta_B^0} = \mu_B (B_x g_{x0} +i B_z g_{z0})$.  
To derive the MF wave functions,\cite{Klinovaja2012} we first linearize the spectrum around the Fermi points $k_z=0$ (interior branch of the spectrum) and $k_z=\pm k_F^s$ (exterior branch of the spectrum) and express the fermionic operators in terms of slowly varying left and right movers ${L}_{\pm}$ and ${R}_{\pm}$ where the indices $\pm$ label the two pseudospin directions for a quantization axis pointing along the SOI induced by the electric field $E_x$. As a result, the Hamiltonian splits into two independent parts. The exterior branch is described by
\begin{equation}
\mathscr{H}^{e} =  i \hbar v_F^s  \eta_0 \nu_z \partial_z + \frac{1}{2}\left[i \Delta_{sc}(\eta_x + i \eta_y) \nu_y  + \text{H.c.}\right],\label{eq:HamouterlabstrongSOI}
\end{equation}
which is written in the basis $({L}_+, {R}_-, {L}_+^{\dagger}, {R}_-^{\dagger} )$.
Here, the Pauli matrices $\eta_i$ ($\nu_i$), $i=0,x,y,z$, act in particle-hole (left and right mover) subspace.
The interior branch is described by
\begin{align}
\mathscr{H}^{i} =  & - i \hbar v_F^s \eta_0 \nu_z\partial_z + \frac{1}{2}\left[i \Delta_{sc}(\eta_x + i \eta_y) \nu_y  + \text{H.c.}\right]\nonumber\\
&- \Delta_Z^0 (\cos \vartheta_B^0 \eta_0 \nu_y + \sin \vartheta_B^0 \eta_z \nu_x ),\label{eq:HaminnerlabstrongSOI}
\end{align}
which is given in the basis $({R}_+, {L}_-, {R}_+^{\dagger}, {L}_-^{\dagger})$. As shown before, a localized zero energy state, MF, exists in the topological phase $\Delta_Z^0 >|\Delta_{sc}|$. The associated  MF wave function is a sum of two contributions,\cite{Klinovaja2012}  $\hat{\Phi}_{s}(z)  = \hat{\Phi}_{s}^e(z) + \hat{\Phi}_{s}^{i}(z)$, originating from the exterior and interior branches (for an explicit expression see Appendix \ref{app:wavefunctions_strong}), which are of the form
\begin{equation}
\hat{\Phi}_{s}^e(z) \propto e^{- z/\xi_s^e}, \qquad
\hat{\Phi}_{s}^i(z) \propto e^{- z/\xi_s^i}, \label{eq:decayingcomps_strongSOI}
\end{equation}
with the localization lengths given by
\begin{align}
\xi_s^e &
= \frac{\alpha_{\text{eff}} |E_x|}{|\Delta_{sc}|},\label{eq:decaystrongSOI_e} \ \ \ 
\xi_s^i 
=\frac{\alpha_{\text{eff}} |E_x|}{|\Delta_{-}|}. 
\end{align}
Both $\xi_s^e$ and $\xi_s^i$ depend linearly on the magnitude of the applied electric field $E_x$, thus weaker fields result in smaller localization lengths. 
Furthermore, $\xi_s^i$ shows an implicit dependence on the magnitude and direction of $\bm{B}$ in the denominator via $\Delta_{-} = |\Delta_{sc}|-\Delta_Z^0$.
The localization length diverges when $\Delta_Z^0$ approaches $|\Delta_{sc}|$, which happens when the topological gap closes and the system becomes gapless.
In Fig.~\ref{fig:localizationlengths_strongSOI_Bangledep}, we plot $\xi_s^e$ and $\xi_s^i$ as functions of the angle $\theta$ enclosed by $\bm{B}$ and the $x$ axis (see Fig.~\ref{fig:wiresetup}).  
We see that as soon as $\bm{B}$ has a nonzero component parallel to the NW, $\xi_s^i$ increases until it diverges at the point where the topological gap closes and the system goes into the topologically trivial phase. 
This effect roots in the strong anisotropy of the $g$ factor in $\Delta_Z^0$ which leads to a quick closing of the topological gap as soon as $\theta$ deviates from zero. 
Furthermore, increasing the magnitude of $\bm{B}$ while still being in the strong SOI regime, such that $|\Delta_{-}|> |\Delta_{sc}|$, i.e.\ $\Delta_Z^0\geq 2|\Delta_{sc}|$, results in $\xi_s^i<\xi_s^e$ for a certain range of $\theta$.  Thus, in this regime, the localization length of the MF wave functions, $\xi={\rm{ max}} \{\xi_s^i, \xi_s^e\}$, is independent of the magnetic field.
%
%
%
%
%
%
%
%
\subsection{Weak SOI  \label{sec:derivationwavefunct_weakSOI}}
\begin{figure}[tbp]
\includegraphics[width=\columnwidth]{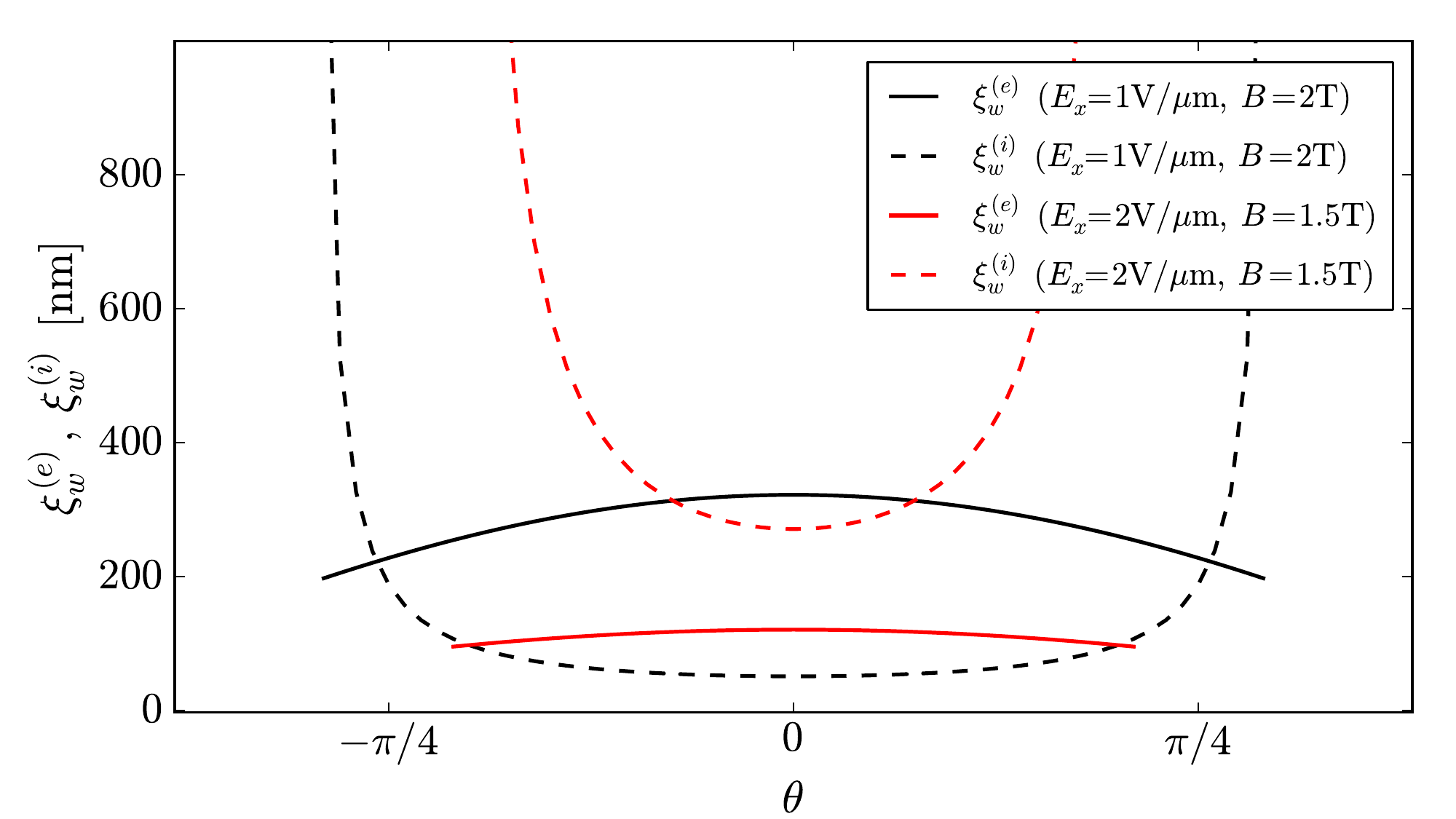}
\caption{Localization lengths $\xi_w^{(e)}$ and $\xi_w^{(i)}$ as functions of the angle $\theta$ defined by $\bm{B} =B(\cos\theta,0,\sin\theta)$, for different magnitudes of $E_x$ and $B$. 
Here we use the combinations $E_x = 1 \mbox{ V/$\mu$m}$ and $B = 2 \mbox{ T}$ (black), $E_x = 2 \mbox{ V/$\mu$m}$ and $B = 1.5 \mbox{ T}$ (red). 
As soon as $\theta$ deviates from $0$, $\xi_w^{(i)}$ increases until its value diverges when the topological gap closes. 
In contrast to $\xi_s^e$ (see Fig.~\ref{fig:localizationlengths_strongSOI_Bangledep}), $\xi_w^{(e)}$ now shows a dependence on $\theta$. 
Here, we use the same values for the NW parameters and superconducting pairing parameter as in Fig.~\ref{fig:localizationlengths_strongSOI_Bangledep}. 
\label{fig:localizationlengths_weakSOI_Bangledep}}
\end{figure}

Next, we focus on the weak SOI regime in which $\Delta_{so}\ll\Delta_Z$. We find that $\Delta_Z\sim \Delta_Z^0$ is still a good approximation for magnetic fields up to $B\sim5\mbox{T}$. This allows us to simplify Eq.~(\ref{eq:kFermi_general}). 
As a result, the Fermi wave number is given by $k_F^w \approx \sqrt{2 m_{\text{eff}} \Delta_Z^0}/\hbar$, 
and the associated Fermi velocity reads $v_F^w = \sqrt{2 \Delta_Z^0/m_{\text{eff}}}$.
By treating the SOI as a weak perturbation, we find the eigenstates of the particle Hamiltonian around the Fermi points $k_F^w$ and linearize the particle-hole Hamiltonian in the basis constructed of these states. We find 
\begin{equation}
\mathscr{H}^{(e)} = - i \hbar v_F^w \eta_0 \nu_z  \partial_z + \frac{1}{2}\left[\bar{\Delta}_{sc}^{*} i e^{-i \vartheta_B^0} (\eta_x + i \eta_y)\nu_y+ \text{H.c.}\right],  \label{eq:HamweakSOI}
\end{equation}
which is represented in the basis $(R, L, R^{\dagger}, L^{\dagger})$, with $R\,(L)$ being the right mover (left mover) in the weak SOI regime. 
As already found in Ref.~[\onlinecite{Klinovaja2012}], the effective coupling due to superconductivity is suppressed by a factor $k_{so}/k_F^w\ll1$, resulting in an effective superconducting coupling term $|\bar{\Delta}_{sc}| = 2 |\Delta_{sc}| k_{so}/k_F^w$.
This can be understood from the fact that without the SOI, the pseudospins of the states at the Fermi points are perfectly aligned and only the weakly perturbing SOI term tilts them into a slightly nonparallel configuration that enables a superconducting pairing.
The localized MF wave function is again a sum of two contributions, $\hat{\Phi}_{w}(z)  = \hat{\Phi}_{w}^{(e)}(z) + \hat{\Phi}_{w}^{(i)}(z)$ (for an explicit expression see Appendix \ref{app:wavefunctions_weak}), which are of the form
\begin{equation}
\hat{\Phi}_{w}^{(e)}(z) \propto e^{- z/\xi_w^{(e)}}, \qquad
\hat{\Phi}_{w}^{(i)}(z) \propto e^{- z/\xi_w^{(i)}},  \label{eq:decayingcomps_weakSOI}
\end{equation} 
with localization lengths given by
\begin{align}
\xi_w^{(e)} &
 =  \sqrt{\frac{2 \Delta_Z^0}{m_{\text{eff}}}}\frac{\hbar}{|\bar{\Delta}_{sc}|}
 = \frac{\Delta_Z^0}{ |\Delta_{sc}|} \frac{\hbar^2}{m_{\text{eff}} \alpha_{\text{eff}} |E_x|},\nonumber\\
\xi_w^{(i)}
& = \frac{\alpha_{\text{eff}} |E_x|}{|\Delta_{-}|}. \label{eq:decayweakSOI_i}
\end{align}
%
%
%
%
%
%
%
%
%
We see that the localization lengths depend quite differently on the strength of the SOI determined by $E_x$, $\xi_w^{(e)}\propto 1/|E_x|$ and $\xi_w^{(i)}\propto |E_x|$. 
Furthermore, in contrast to the strong SOI regime, both localization lengths depend on the magnitude and direction of $\bm{B}$.
In Fig.~\ref{fig:localizationlengths_weakSOI_Bangledep}, we plot $\xi_w^{(e)}$ and $\xi_w^{(i)}$ as functions of the angle $\theta$ enclosed by $\bm{B}$ and the $x$ axis, for various combinations of $E_x$ and $B$. 
We observe that the dependence of $\xi_w^{(e)}$ on $\theta$ is much weaker than the $\theta$ dependence of $\xi_w^{(i)}$, e.g.,  $\xi_w^{(e)}$ does not diverge when the topological gap closes. 
Furthermore, depending on the relative magnitude of the fields, we can find both, $\xi_w^{(e)}>\xi_w^{(i)}$ and $\xi_w^{(e)}<\xi_w^{(i)}$. 
%
%
%
%
%
\subsection{Optimal experimental regime}

\begin{figure}[tbp]
\includegraphics[width=\columnwidth]{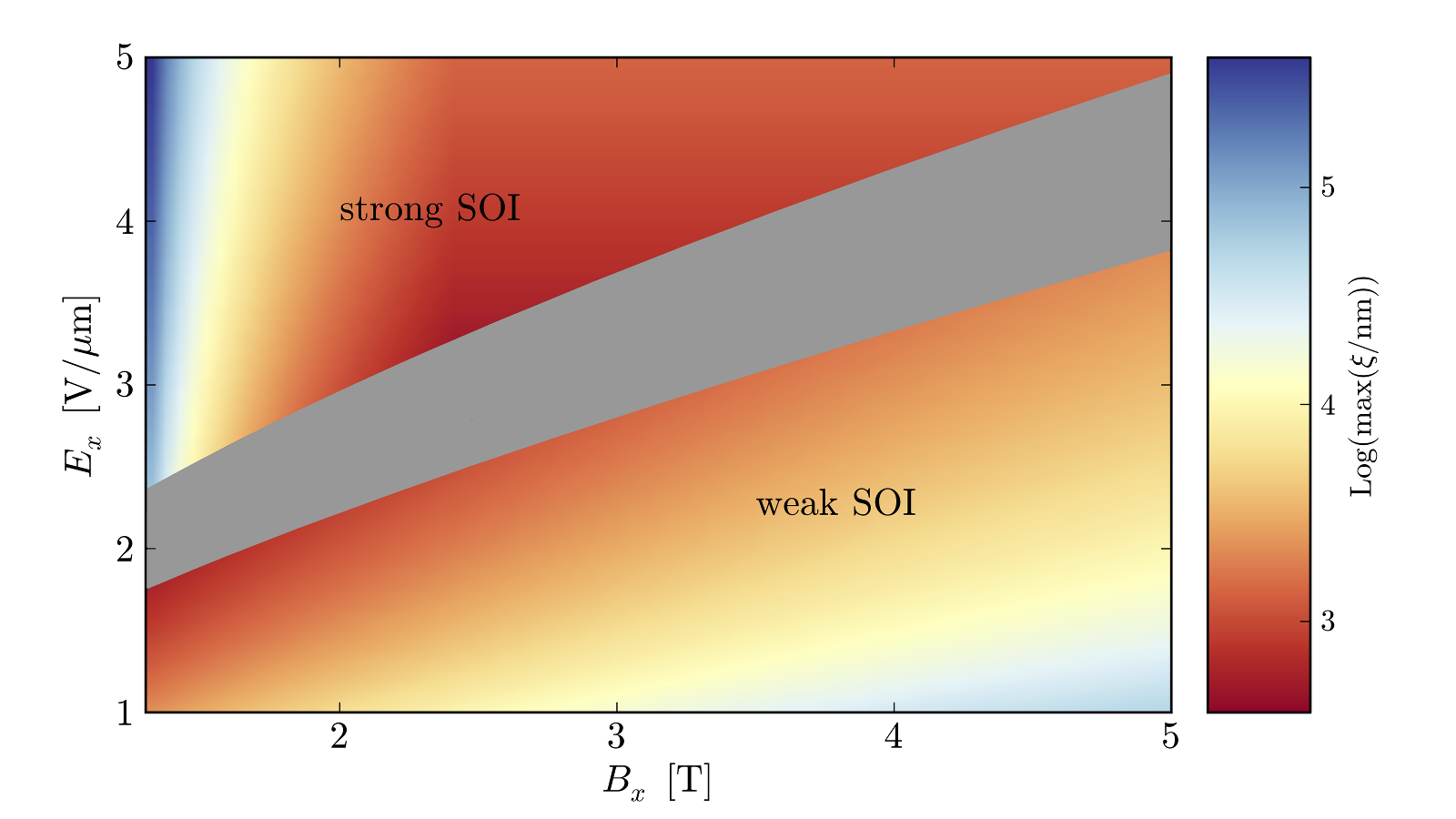}
\caption{Logarithm (color coded) of the dominating localization length, $\xi =\max\{\xi_s^e,\xi_s^i\}$ and $\xi_w^{(e)}$, in the strong and weak SOI regime, respectively, as functions of the applied fields $E_x$ and $\bm{B}$. 
For simplicity we restrict ourselves to $\bm{B}=(B_x,0,0)$. 
The diagonal gray area approximately denotes the transitional regime between regions of strong and weak SOI. 
Here, we use the same values for the NW parameters and superconducting pairing parameter as in Fig.~\ref{fig:localizationlengths_strongSOI_Bangledep}. 
\label{fig:localizationlengths_bothphases}}
\end{figure}

As shown above, the magnitude and direction of the applied fields determine the  localization lengths of the MF wave functions; see Figs.~\ref{fig:localizationlengths_strongSOI_Bangledep} and \ref{fig:localizationlengths_weakSOI_Bangledep}. 
In experiments, it is crucial to tune the applied fields such that obtained MFs are well separated and the localization lengths are as short as possible.
To identify the optimal field regime, we display the logarithm of the maximal localization length for the given magnitudes of the applied fields in Fig.~\ref{fig:localizationlengths_bothphases}. 
In the weak SOI regime, we 
furthermore have to take into account that the part of the wave function decaying with the localization length $\xi_s^{(i)}$ is additionally suppressed by a factor $k_{so}/k_F^w\ll1$ and thus can be neglected. 
To simplify the analysis, we fix the direction of the magnetic field to be perpendicular to the NW, $\bm{B} = (B_x,0,0)$. Here, we are motivated by the fact that this configuration corresponds to the shortest localization lengths in the strong SOI regime.
The range of $B_x$ is chosen  such that we remain in the topological regime throughout. 
We see that when applying large $E_x$ while keeping $B_x$  small or when applying large $B_x$ while keeping $E_x$ small, the localization lengths are not the shortest possible. Hence it is most favorable to choose an intermediate regime in which both fields take rather moderate values and the Zeeman energy and the SOI energy are comparable with each other.

\FloatBarrier
%
%
%
%
%
\section{Finite Nanowires: Hybridized Majorana fermions \label{sec:Energyoscillations}}
So far we have focused on a semi-infinite NW that, when being brought into the topological phase, hosts a zero-energy MF at the end.
However, in any realistic system, the NWs are of finite length $L$ and host two MFs: one MF at each end. These MFs could overlap and hybridize if their localization lengths are comparable with the NW length. This results in the emergence  of an ordinary fermion which, in general, possesses a nonzero energy.\cite{DasSarma2012, Rainis2013} In this section, we examine the dependence of this fermionic energy on the magnitude of the applied fields in the strong and weak SOI regime.
We assume that the NW stretches from $z=0$ to $z=L$ and search for hybridized wave functions that satisfy vanishing boundary conditions at both ends of the NW. We note that we focus here on the direct overlap between two MF wave functions and neglect a possible hybridization mediated by bulk superconducting states. \cite{Zyuzin2013}
%
%
%
%
%
\subsection{Strong SOI\label{sec:oscillatingenergy_strongSOI}}
\begin{figure}[t!]
\includegraphics[width=\columnwidth]{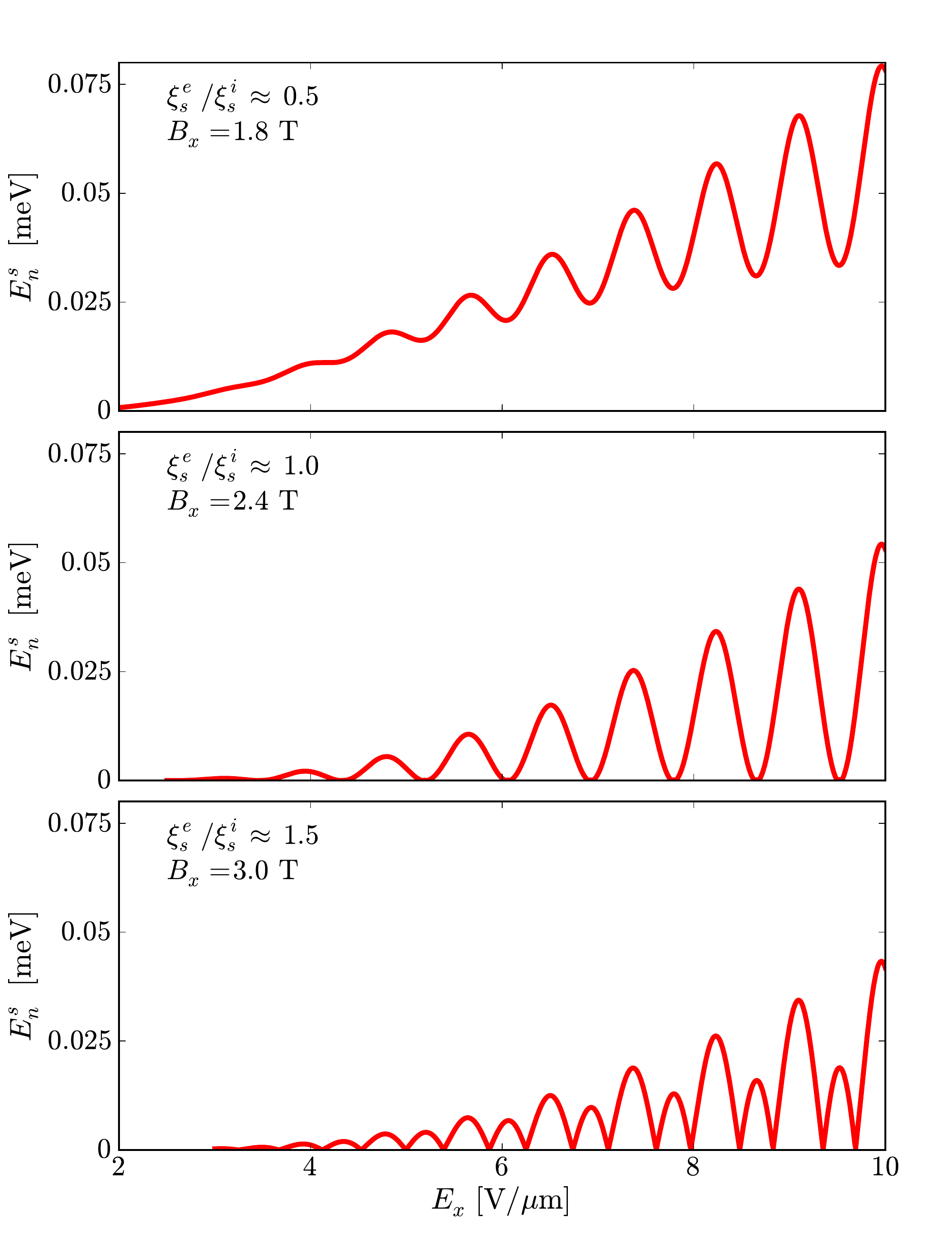}
\caption{The fermion energy $E_n^s$ in the strong SOI regime as a function of the electric field $E_x$ for different magnitudes of the applied magnetic field $\bm{B} = (B_x,0,0)$; see the insets. 
We consider three different ratios of the localization lengths $\xi_s^e/\xi_s^i$. If the contribution of the interior branches is dominant, $\xi_s^i > \xi_s^e$, $E_n^s$ shows an increasing offset from zero with weak superimposed oscillations on  top of it.  If the contribution of the exterior branches is dominant, $\xi_s^e > \xi_s^i$, $E_n^s$ shows an increasing offset from zero with strong superimposed oscillations on the top of it such that  $E_n^s$ periodically returns to zero.
Here, we assume a superconducting pairing potential $\Delta_{sc} = 200 \mbox{ $\mu$eV}$ and use the NW parameters $R=7.5\mbox{ nm}$, $\Delta = 23\mbox{ meV}$, and $L = 0.7\mbox{ $\mu$m}$.
\label{fig:oscillatingenergy}}
\end{figure}
In this section, we explore the energy of hybridized MFs in the strong SOI regime. First, we solve the Schr\"odinger equations $\mathscr{H}^{e} \phi^e = E_n^s \phi^e$ and  $\mathscr{H}^{i} \phi^i = E_n^s \phi^i$ for  arbitrary $E_n^s>0$ (for the Hamiltonian densities see Sec.~\ref{sec:derivationwavefunct_strongSOI}). 
Since the NW is of finite length, both the decaying and growing eigenfunctions are normalizable and may contribute to the final hybridized wave function. 
We obtain a set of eight eigenfunctions $\phi^{e}_j$ and $\phi^{i}_j$, $j=1,\ldots,4$, where the $\phi^{e}_j$ show an oscillatory behavior proportional to  $e^{\pm i k_F^s z}$.
We search for a nontrivial linear combination of these eigenfunctions, $\Phi_{\text{hyb}}^s (z) = \sum_{j} (a^e_j \phi^{e}_j +a^i_j \phi^{i}_j)$, that satisfies the boundary conditions $\Phi_{\text{hyb}}^s (z=0)=\Phi_{\text{hyb}}^s (z=L)=0$. 
In the considered regime, $E_n^s$ is given by a quite involved implicit equation which we omit displaying here. 

We provide numerical results for several specific sets of applied fields and plot $E_n^s$ as a function of $E_x$ in Fig.~\ref{fig:oscillatingenergy}. For all field configurations, we observe an oscillatory behavior of $E_n^s$ with increasing amplitude as $E_x$ increases. In addition, depending on the magnitude of the applied magnetic field, the curves may show a nonzero offset that increases with $E_x$. This  feature is the most pronounced for small magnetic fields close to the point where the topological gap closes. 

These results can be explained easily when remembering the zero energy MF wave function which is a linear combination of oscillating (exterior branch, $\xi_s^e$) and nonoscillating (interior branch, $\xi_s^i$)  wave functions (see Sec.~\ref{sec:derivationwavefunct_strongSOI}). This result remains valid for very small energies $E_n^s$, thus the lengthscales governing the growth and decay of the eigenstates $\phi^{e,i}_j$ are comparable to  $\xi_s^e$ and $\xi_s^i$. 
If the localization length is set by the nonoscillating interior branch part, $\xi_s^e/\xi_s^i\ll1$ (the respective ratios are included in each plot in Fig.~\ref{fig:oscillatingenergy}), $E_n^s$  monotonically splits away from zero with superimposed weak oscillations. On the contrary, if the localization length is set by the strongly oscillating exterior branch part, $\xi_s^e/\xi_s^i\geq1$, $E_n^s$ oscillates strongly and even goes back to zero.
Thus, depending on the ratio of the localization lengths $\xi_s^e$ and $\xi_s^i$ which can be tuned by changing the magnitude or the direction of the applied magnetic field with respect to the NW (see Fig.~\ref{fig:localizationlengths_strongSOI_Bangledep}), the offset of  $E_n^s$ can be tuned.
The period of the superimposed oscillations is independent of the magnitude of $B_x$ since the $\phi^{e}_j$ cause oscillations with a period $\delta E_x = \hbar^2 \pi/(m_{\text{eff}} \alpha_{\text{eff}} L)$. Using the latter relation, the strength of the SOI can be determined from the oscillation period.

%
%
%
%
%
%
%
\subsection{Weak SOI \label{sec:oscillatingenergy_weakSOI}}
\begin{figure}[t!]
\includegraphics[width=\columnwidth]{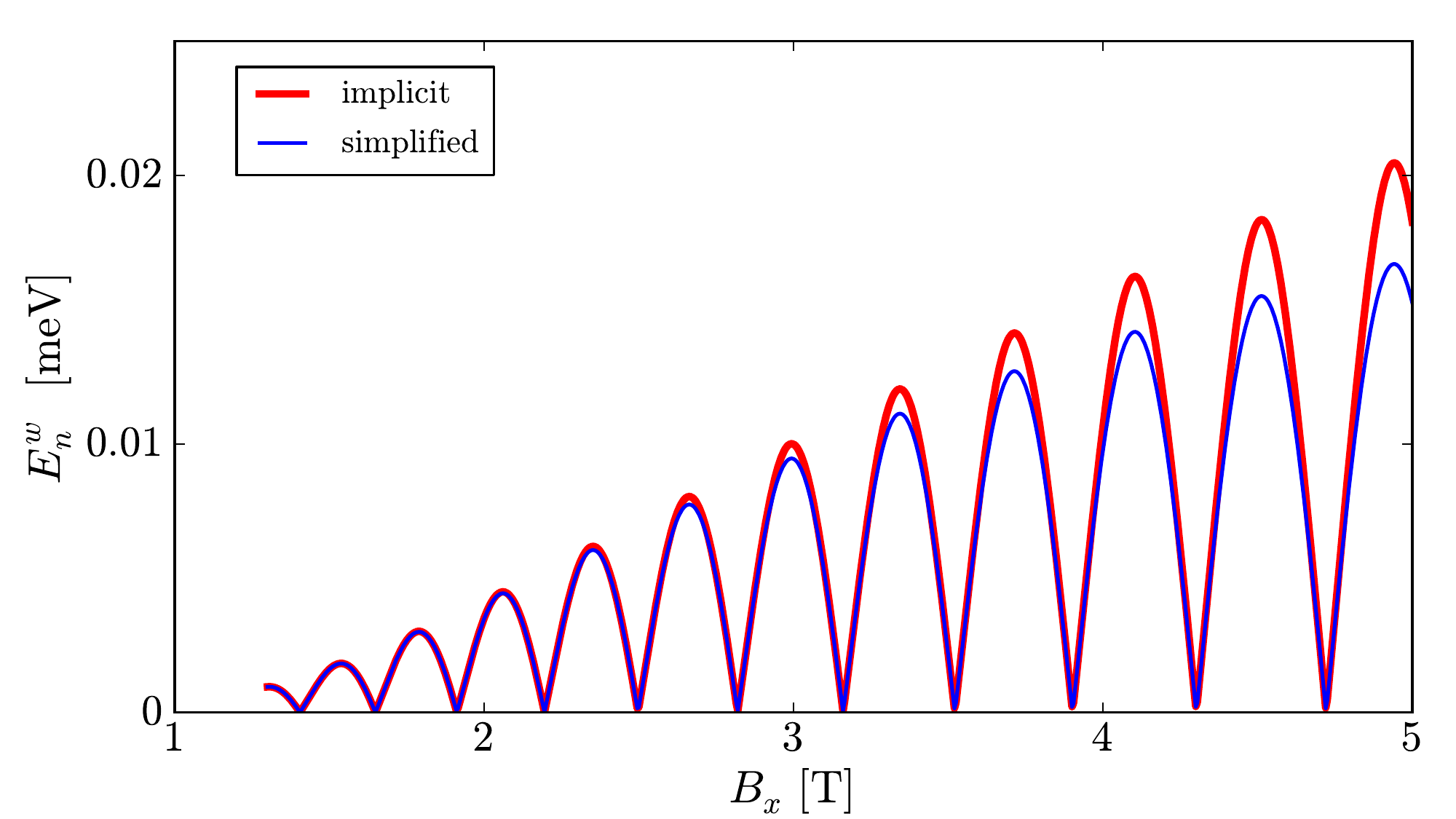}
\caption{The oscillating energy $E_n^w$ as a function of $\bm{B} = (B_x ,0,0)$. 
We display both the results from solving the implicit Eq.~(\ref{eq:EnimplicitweakSOI}) numerically (red) and the simplified solution given in Eq.~(\ref{eq:EnimplicitweakSOIsimp}) (blue) and find good agreement. 
We assume a weak electric field $E_x = 0.5\mbox{ V/$\mu$m}$ and a proximity induced superconducting pairing potential $\Delta_{sc} = 200 \mbox{ $\mu$eV}$.
We use the NW parameters $R=7.5\mbox{ nm}$, $\Delta = 23\mbox{ meV}$, and $L = 2\mbox{ $\mu$m}$.
\label{fig:oscillatingenergy_weakSOI}}
\end{figure}

In the weak SOI regime, we apply the same procedure as described in Sec.~\ref{sec:oscillatingenergy_strongSOI}.
Here, we employ the Hamiltonian $\mathscr{H}^{(e)}$ given in Eq.~(\ref{eq:HamweakSOI}) and solve the Schr\"odinger equation $\mathscr{H}^{(e)}\phi^{(e)} = E_n^w \phi^{(e)}$ for an arbitrary $E_n^w>0$. 
The four eigenstates $\phi^{(e)}_j$, $j=1,\ldots,4$, are combined into a nontrivial linear combination, $\Phi_{\text{hyb}}^w (z) = \sum_{j} b^{(e)}_j \phi^{(e)}_j$,
that satisfies the boundary conditions $\Phi_{\text{hyb}}^w(z=0)=\Phi_{\text{hyb}}^w (z=L)=0$. This leads to an implicit condition for $E_n^w$, 
\begin{equation}
\left[\frac{\sqrt{\bar{\Delta}_{sc}^2-E_n^2}}{\bar{\Delta}_{sc}}\right]^2=\frac{2 \sinh ^2\left[\frac{L \sqrt{\bar{\Delta}_{sc}^2-E_n^2}}{v_F^w \hbar }\right]}{\cosh \left[\frac{2 L \sqrt{\bar{\Delta}_{sc}^2-E_n^2}}{v_F^w \hbar }\right]-\cos (2 k_F^w L)}.  \label{eq:EnimplicitweakSOI}
\end{equation}
This implicit equation can be transformed to an explicit relation for $E_n^w$ by assuming that $E_n^w\ll\bar{\Delta}_{sc}$, 
\begin{equation}
E_n^w \approx \bar{\Delta}_{sc} |\sin(k_F^w L)| e^{-L/\xi_w^{(e)}},  \label{eq:EnimplicitweakSOIsimp}
\end{equation}
which shows an oscillatory behavior of $E_n^w$ as a function of $k_F^w$, where the latter is a function of $\bm{B}$. 
In Fig.~\ref{fig:oscillatingenergy_weakSOI}, we plot $E_n^w$ as a function of $\bm{B} = (B_x ,0,0)$ for a weak electric field $E_x$ where we obtained $E_n^w$ once by solving Eq.~(\ref{eq:EnimplicitweakSOI}) numerically  and once by using the explicit relation given in Eq.~(\ref{eq:EnimplicitweakSOIsimp}).
Both results agree well, especially for small $B_x$.  
We see that $E_n^w$ oscillates and the energy of the fermion composed of two overlapping MFs periodically comes back to zero.\cite{DasSarma2012,Rainis2013}
The periodicity of the oscillation is given by 
\begin{equation}
\delta\Delta_Z^0  \approx \frac{\pi \hbar}{L}\sqrt{\frac{{2 \Delta_Z^0}}{{m_{\text{eff}}}}}  , 
\end{equation}
and depends on the $g$ factor and hence on the direction of the magnetic field. 
As a result, the change in the period as a function of the magnetic field direction could be used to measure the $g$ factor anisotropy. 

Changing the strength of the applied electric field $E_x$ does not affect the period of the oscillation, however, examining Eq.~(\ref{eq:EnimplicitweakSOIsimp}) in more  detail shows that a stronger field $E_x$ yields a smaller amplitude of the splitting.

%
%
%
%
%
%
%
\section{Conclusion \label{sec:conclusion}}
In this work, we utilized a concrete microscopic model to study MFs confined to Ge/Si core/shell NWs. To this end, we derived an effective 1D lowest subband hole Hamiltonian, which also includes the superconducting pairing and takes into account specifics of Ge/Si core/shell NWs such as the $g$ factor anisotropy and the electric field dependence of the induced Rashba SOI.
We have determined the MF localization lengths in the strong and in the weak SOI regime and examined their dependence on the direction of the magnetic field with respect to the NW. 
In general, we found that intermediate magnitudes of electric and magnetic fields lead to the shortest localization lengths of the MF wave functions. 
Additionally, we examined finite NWs, where two MFs localized at the NW ends overlap and form an ordinary fermion. 
This hybridization results in a fermion whose energy oscillates as a function of the electric field (magnetic field) in the strong (weak) SOI regime. The possibility to control the overlap of MFs by tuning only electric fields could be used to perform topologically nonprotected operations on Majorana fermion that are necessary to realize universal quantum computation.\cite{Alicea2012}

From our results we conclude that Ge/Si core/shell NWs are a promising system regarding the emergence of MFs due to the high control over the SOI strength. In addition, we note that these NWs are also excellent candidates for parafermion setups which require even stronger SOI and substantial electron-electron interactions. \cite{Oreg2014, Klinovaja2014b,Klinovaja2014c}
In addition, the lowest energy subband Hamiltonian derived in this work can provide the basis for further investigation of arrays of Ge/Si NWs. Such arrays could also be used to study quantum Hall effect,  \cite{Teo2014,Klinovaja2013d, Neupert2014, Meng2014, Klinovaja2014e} topological superconductor,\cite{Seroussi2014,Sagi2014} and quantum spin Hall effect\cite{Klinovaja2014d} in the anisotropic limit.

%
%
%
%
%
\begin{acknowledgments}
We thank Christoph Kloeffel for helpful discussions. We acknowledge support from the Swiss NSF, NCCR QSIT, through the EC FP7-ICT initiative under project SiSPIN No. 323841, and the Harvard Quantum Optical Center.
\end{acknowledgments}

%
%
%
%
%
\begin{appendix}
\begin{widetext}
\section{Particle-hole coupling Hamiltonian \label{app:Hamiltonian}}
In this section, we display the effective Bogoliubov-de Gennes Hamiltonian for holes in the lowest-energy subband. To allow for a direct comparison with previous results,\cite{Klinovaja2012} we rotate the lowest-energy basis $\Psi_{ph}$ [introduced above Eq.~(\ref{eq:Hamsupercondeff1D})] such that the spin quantization axis lies parallel to the applied electric field $E_x$ and the new particle-hole basis reads  $\tilde{\Psi}_{ph} = (\Psi_{+},\Psi_{-}, \Psi_{+}^{\dagger},\Psi_{-}^{\dagger})$, where the $\pm$ denotes the pseudospin of the SOI split subband. The Hamiltonian is given by $\tilde{H}_{ph} = \frac{1}{2}\int \tilde{\Psi}_{ph}^{\dagger}\tilde{\mathscr{H}}_{ph}\tilde{\Psi}_{ph}$ with Hamiltonian density 

\begin{equation}
 \tilde{\mathscr{H}}_{ph} =\left(\begin{array}{cccc}
    \frac{\hbar^2 k_z^2}{2 m_{\text{eff}}} -\mu + E_x \alpha_{\text{eff}} k_z & i \Delta_Z e^{i \vartheta_B}	& 0	&\Delta_{sc}\\
    -i \Delta_Z e^{-i \vartheta_B}& \frac{\hbar^2 k_z^2}{2 m_{\text{eff}}} -\mu - E_x \alpha_{\text{eff}} k_z	&-\Delta_{sc}	& 0\\
    0	& -\Delta_{sc}^{*} & -\frac{\hbar^2 k_z^2}{2 m_{\text{eff}}} +\mu + E_x \alpha_{\text{eff}} k_z	& i \Delta_Z e^{-i \vartheta_B}\\
  \Delta_{sc}^{*} 	& 0	& -i \Delta_Z e^{i \vartheta_B}	& -\frac{\hbar^2 k_z^2}{2 m_{\text{eff}}} +\mu - E_x \alpha_{\text{eff}} k_z
\end{array}\right). \label{eq:HpartholeRotSOI_app}
\end{equation}
Here, we used the abbreviation $\Delta_Z e^{i \vartheta_B} = \mu_B (B_x g_x +i B_z g_z)$. 
A similar Hamiltonian was used in Ref.~[\onlinecite{Klinovaja2012}] to derive MF wave functions in NWs with proximity-induced superconductivity. Note that our model additionally includes a complex superconducting pairing potential and a Zeeman term reflecting the anisotropy of the $g$ factor of the NW. 

%
%
%
%
%
\section{Wave functions \label{app:wavefunctions}}
Here we display the explicit form of the MF wave functions in both the strong and weak SOI regime. 
\subsection{Strong SOI\label{app:wavefunctions_strong}}
In the strong SOI regime, the MF wave function introduced above Eq.~(\ref{eq:decayingcomps_strongSOI}) in Sec.~\ref{sec:derivationwavefunct_strongSOI} is given by
\begin{align}
\hat{\Phi}_{s}(z)  =& e^{-i \pi/4}  e^{- z/\xi_s^e}\left(
\begin{array}{c}
i  e^{i\vartheta_B^0/2}e^{i k_F^s z}\\
- i e^{-i\vartheta_B^0/2}e^{-i k_F^s z}\\
  e^{-i\vartheta_B^0/2}e^{-i k_F^s z}\\
- e^{i\vartheta_B^0/2}e^{i k_F^s z}
\end{array}
\right)+e^{-i \pi/4}  e^{-z/\xi_s^i} \left(
\begin{array}{c}
-i e^{i\vartheta_B^0/2}\\                              
ie^{-i\vartheta_B^0/2}\\
-e^{-i\vartheta_B^0/2}\\
 e^{i\vartheta_B^0/2}
\end{array}
\right) ,   \label{eq:Majoranamode_strongSOI_right_app}
\end{align}
where $\hat{\Phi}_{s}(z)$ is written in the basis $\tilde{\Psi}_{ph}$ (see Appendix  \ref{app:Hamiltonian}).

\subsection{Weak SOI\label{app:wavefunctions_weak}}
In the weak SOI regime, the MF wave function introduced above Eq.~(\ref{eq:decayingcomps_weakSOI}) in Sec.~\ref{sec:derivationwavefunct_weakSOI} is given by
\begin{align}
\hat{\Phi}_{w}(z) =& \frac{1}{\sqrt{2}}e^{-z/\xi_w^{(e)}}\left(\begin{array}{c}
e^{i\vartheta_B^0/2}
\left[\left(1-\frac{k_{so}}{k_F^w}\right)\left(1-i\frac{k_{so}}{k_F^w}\right)e^{i {k_F^w} z}-\left(1+\frac{k_{so}}{{k_F^w}}\right)\left(1+i\frac{k_{so}}{{k_F^w}}\right)e^{-i {k_F^w} z}\right]\\
e^{-i\vartheta_B^0/2}
\left[\left(1+\frac{k_{so}}{{k_F^w}}\right)\left(i+\frac{k_{so}}{{k_F^w}}\right)e^{i {k_F^w} z}-\left(1-\frac{k_{so}}{{k_F^w}}\right)\left(i-\frac{k_{so}}{{k_F^w}}\right)e^{-i {k_F^w} z}\right]\\
e^{-i\vartheta_B^0/2}
\left[\left(1-\frac{k_{so}}{{k_F^w}}\right)\left(1+i\frac{k_{so}}{{k_F^w}}\right)e^{-i {k_F^w} z}-\left(1+\frac{k_{so}}{{k_F^w}}\right)\left(1-i\frac{k_{so}}{{k_F^w}}\right)e^{i {k_F^w} z}\right]\\
e^{i\vartheta_B^0/2}
\left[\left(1+\frac{k_{so}}{{k_F^w}}\right)\left(-i+\frac{k_{so}}{{k_F^w}}\right)e^{-i {k_F^w} z}-\left(1-\frac{k_{so}}{{k_F^w}}\right)\left(-i-\frac{k_{so}}{{k_F^w}}\right)e^{i {k_F^w} z}\right]
\end{array} \right)\nonumber\\
&+ 2e^{-z/\xi_w^{(i)}}\frac{k_{so}}{{k_F^w}}e^{-i\pi/4}\left(\begin{array}{c}
-i e^{i\vartheta_B^0/2}\\
i e^{-i\vartheta_B^0/2}\\
-e^{-i\vartheta_B^0/2}\\
e^{i\vartheta_B^0/2}
\end{array} \right), \label{eq:Majoranamode_weakSOI_right_app}
\end{align}
where $\hat{\Phi}_{w}(z)$  is written in the basis $\tilde{\Psi}_{ph}$ (see Appendix \ref{app:Hamiltonian}).
For very weak SOI ($k_{so}/k_F^w\ll1)$ the MF wave function simplifies to
\begin{equation}
\hat{\Phi}_{w}(z) \approx  \sqrt{2}   \sin({k_F^w} z)e^{-z/\xi_w^{(e)}} \left(\begin{array}{c}
i e^{i\vartheta_B^0/2}  \\
- e^{-i\vartheta_B^0/2}  \\
-i e^{-i\vartheta_B^0/2} \\
- e^{i\vartheta_B^0/2}  
\end{array} \right). 
\end{equation}

\end{widetext}

\end{appendix}
%
%
%
%
%
%

%

\end{document}